# Elastic Time Reversal Mirror Experiment in a Mesoscopic Natural Medium at the Low Noise Underground Laboratory of Rustrel, France

Stéphane Gaffet[1,2], Tony Monfret[2], Guy Sénéchal[3], Dominique Rousset[3], Hermann Zeyen[4], Michel Auguste[1], Daniel Boyer[1] and Alain Cavaillou[1]

[1] LSBB     Laboratoire Souterrain à Bas Bruit, UNS/CNRS/OCA, la grande combe, 84400 Rustrel, France (http://lsbb.oca.eu)
[2] GEOAZUR     UMR 6526, UNS/CNRS/IRD/OCA, Sophia-Antipolis, France
[3] MIGP     Modélisation et Imagerie en Géosciences, UMR 5212 UPPA/CNRS/TOTAL, Pau, France
[4] IDES     Interaction et Dynamique des Environnements de surface, UMR 8148 UPS/CNRS, Orsay, France

## ABSTRACT

A seismic time reversal experiment based on **T**ime **R**eversal **M**irror (TRM) technique was conducted in the mesoscopically scaled medium at the LSBB Laboratory, France. Two sets of 50 Hz geophones were distributed at one meter intervals in two horizontal and parallel galleries 100 m apart, buried 250 m below the surface. The shot source used was a 4 kg sledgehammer. Analysis shows that elastic seismic energy is refocused in space and time to the shot locations with good accuracy. The refocusing ability of seismic energy to the shot locations is roughly achieved for the direct field, and with excellent quality, for the early and later coda. Hyper-focussing is achieved at the shot points as a consequence of the fine scale randomly heterogeneous medium between the galleries. TRM experiment is sensitive to the roughness of the mirror used. Roughness induces a slight experimental discrepancy between recording and re-emitting directions degrading the quality of the reversal process.

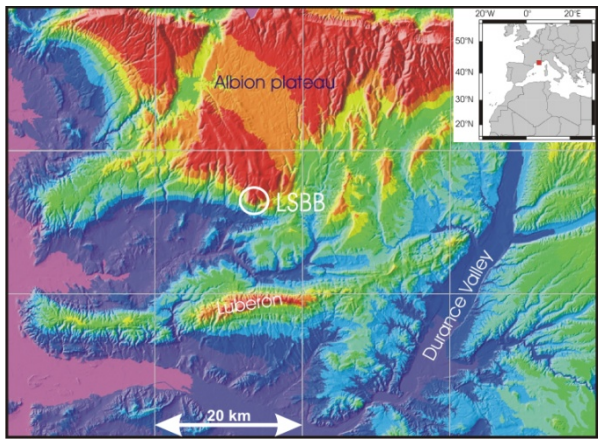 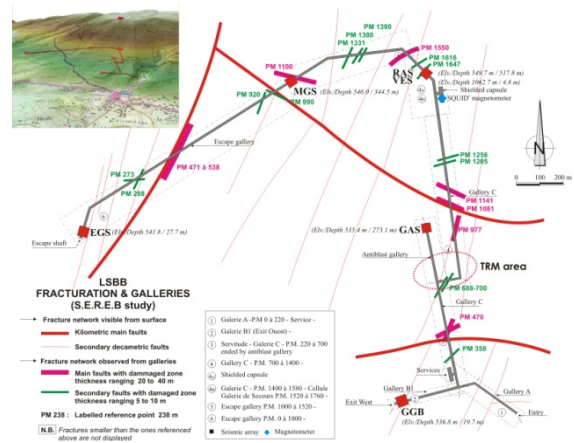

*Figure 1:* Location of LSBB in the southern border of Fontaine-de-Vaucluse aquifer that flows from east to west below the Albion plateau, in the South of France

*Figure 2:* Tunnel topology and fracturing identified from aerial imagery done by the French Institut Géographique National. The dashed circle, indicates the segments of the main gallery (GPR) and the antiblast gallery (GAS) where the TRM experiment was performed.

## INTRODUCTION

The "Laboratoire Souterrain à Bas Bruit" (LSBB or "Underground Low-noise Laboratory") in Rustrel, southern France (Fig. 1), a former command post for the nuclear forces of the French army, was designated in 2009 as a French National Instrumented Infrastructure by the Centre National de la Recherche Scientifique (CNRS), and dedicated to inter-Disciplinary Underground Science and Technology (*i*-DUST). It is currently used to study topics ranging from finding the missing mass in the universe to studying the subsurface storage of $CO_2$ and the aquifer management of water resources and reserves. The facility is horizontally accessible via a 3.7 km long tunnel with an L-shape (Fig. 2). The deepest vault is 518 m below the surface (~1500-meter water equivalent, m.w.e.). The laboratory spans a surface of almost 100 acres, in an area named "la Grande Montagne". All galleries and vaults have a power supply, phone, GPS time, and internet capability with fiber optic communication cables, and are connected to two huts on the surface. The mountain "la Grande Montagne" shields instruments from many noise sources. Moreover, the LSBB is located within the regional natural park of the Luberon, with few anthropogenic perturbations.

Following Larose et al (2004), in a geophysical context, the mesoscopic nature of seismic rays allows them to sample several hundreds of meters. The specific topology of the galleries of LSSB is very conducive, at mesoscopic scales, for the Time Reversal Mirror processing technique. Such an experiment was conducted between the LSSB "anti-blast" (GAS) gallery and the main (GPR) gallery. The two galleries are separated by 100 m of fractured and porous carbonate, partially fluid-saturated rocks (red dashed circled area in Fig 2). The TRM is a physical process which focuses wave energy scattered in space and time, back to the wave source (see Fink and Prada, 2001, for a general review). It does so by retransmitting the whole transmitted scattered wave field back into the medium after time reversal. Knowledge of the propagating medium is not necessary. Lab experiments have shown that the convergence to the source is enhanced as the inhomogeneity of the medium increases and scattering becomes more pronounced (Derode et al, 1995; Fink and Prada, 2001). In addition, the broader the frequency band taken into account, the sharper the focussing of the seismic energy on the source region becomes (Derode et al., 1998). The TRM device acts as a spatio-temporal matched filter on the propagation transfer between the array of receivers and the source target.

## 1. TIME REVERSAL MIRROR EXPERIMENT

The Time Reversal Mirror experiment was conducted in a natural medium of mesoscopic scale: 50 m × 100 m of heterogeneous medium with wavelengths roughly between 5 m and 40 m (frequency range: 100 Hz to 1000 Hz). Two horizontal linear arrays of 50 one-component geophones of 50 Hz natural frequency at one meter interval were bolted horizontally to the GAS and GPR gallery walls. These sensors recorded the reflected, the transmitted and the diffracted waves emitted from a sledgehammer shot struck between the sensors.

The experiment consisted of three steps displayed in Fig. 3. Two steps are field experiments; the third step consists of numerical processing:

- **Step 1:** The forward experiment consists in recording at the 50 sensors in the "*mirror*" gallery (namely GPS gallery), the signals emitted by 50 sledgehammer shots impacted in the "*source*" gallery (namely GAS gallery).

- **Step 2:** The backward experiment is performed by inverting the role of the two galleries. The two data sets obtained, related directly to step 1 and step 2, can be considered as empirical Green functions of the medium for each couple source/receiver by assuming the sledgehammer shot to be a spike mono-directional impulse.

- **Step 3:** The Time Reversal Mirror process is realized numerically as described in §1.1 below that allows producing the results presented herein

Another Time Reversal Mirror experiment was performed by considering the reverse geometry, i.e. GAS gallery being the "*mirror*" gallery and GPR gallery being the "*source*" gallery.

### 1.1 Contribution of one point of the Time Reversal Mirror

Given a point source at $x_o$ acting along direction $j$, given the $j^{th}$ component of an arbitrary receiver $x_p$, $x_o$ and $x_p$ being located in the "*source*" gallery (see Fig. 3 top and middle), and considering the $i^{th}$ component of an arbitrary point $y_m$ within the "*mirror*" gallery, the contribution of $y_m$ as part of the Time Reversal Mirror can be formulated as follows.

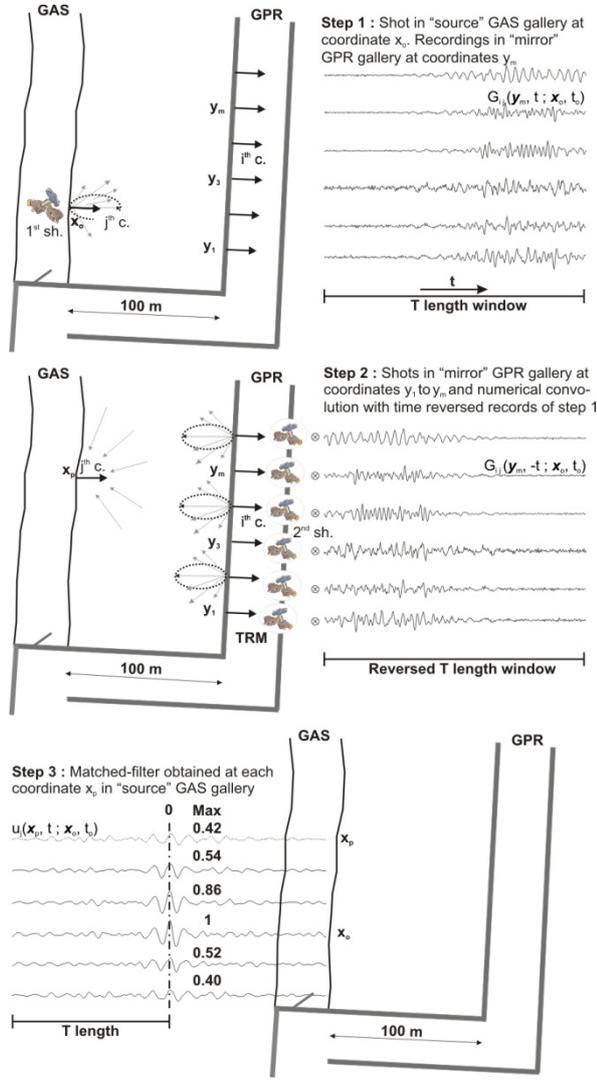

*Figure 3: Step 1 – Generation of elastic "forward" waves with shot in GAS (denoted "source" gallery in the text) recorded in GPR gallery (denoted "mirror" gallery in text). Step 2 – TRM generation of elastic "backward" waves with shots in GPR and recorded in GAS gallery. Step 3 – Matched filter summation of "backward" waves with time reversed "forward" waves in order to realize Eq. 5. Refocussing energy is maximum (Max =1) at origin source point $x_o$.*

The $j^{th}$ component of the motion $u$ at location $x_p$ results from the convolution of two Green functions: *(i)* the Green function $G_{ij}(y_m, t ; x_o, t_o)$ related to the force applied along the $j^{th}$ direction at the source $x_o$ and time $t_o$ recorded along the $i^{th}$ direction at location of the receiver $y_m$ of the "*mirror*" (see step 1 in Fig. 3), and *(ii)* the Green function $G_{ji}(x_p, t ; y_m, t_o+T)$ related to the force applied at point $y_m$ along the $i^{th}$ direction after a delay $T$ and recorded along the $j^{th}$ direction at point $x_p$ (see step 2 in Fig. 3). Hence

$$u_j(x_p, t ; x_o, t_o) = G_{ij}(y_m, -t ; x_o, t_o) \otimes G_{ji}(x_p, t ; y_m, t_o+T) \quad (1)$$

where $\otimes$ is the convolution operator. Experimentally, the first Green function $G_{ij}(y_m, -t ; x_o, t_o)$ in Eq. (1) is obtained using the time reversed ground motion recorded during $T$ seconds in the "*mirror*" gallery at location $y_m$ along the $i^{th}$ component, after the impulse provided by the sledgehammer applied at $t_o$ and located at $x_o$ in the "*source*" gallery roughly along the $j^{th}$ direction (step 1). The second Green function $G_{ji}(x_p, t ; y_m, t_o+T)$ is generated by the sledgehammer shot processed in the "*mirror*" gallery (step 2). This latter shot may be considered as carrying the time reversed direct-field $G_{ij}(y_m, -t ; x_o, t_o)$, back to the "*source*" gallery.

Thus, using time shift property of Green function and assuming that the propagation medium remains unchanged during the experiment; Eq. (1) can be rewritten as follows:

$$u_j(x_p, t ; x_o, t_o) = G_{ij}(y_m, -t ; x_o, t_o) \otimes G_{ji}(x_p, t-T ; y_m, t_o) \quad (2)$$

Applying the reciprocity principle between $y_m$ and $x_p$ (Aki and Richards, 1980), Eq. (2) becomes:

$$u_j(x_p, t ; x_o, t_o) = G_{ij}(y_m, -t ; x_o, t_o) \otimes G_{ij}(y_m, t-T ; x_p, t_o) \quad (3)$$

The reciprocity principle used in Eq. (3) is the most sensitive part of the experiment. It is strongly dependent of the operational conditions since it assumes that *(i)* a sledgehammer impulse seismic radiation can approximate a theoretical point force and *(ii)* the orientation of the sledgehammer impact is well-aligned with the directions of the ground displacements measured in both "*source*" and "*mirror*" galleries. While such conditions can reasonably be accepted for shots and recording onto a planar concrete wall (e.g. as in the GPR gallery), it is poorly controlled for the shots and recordings done onto a naturally corrugated rock wall (e.g. as in the GAS gallery).

Finally, using time delay operator, the component $u_j$ of the motion reconstructed at point $x_p$ by the point $y_m$ of the Time Reversal Mirror, becomes:

$$u_j(x_p, t\,;\,x_o, t_o) = \delta(t-T) \otimes G_{ij}(y_m, -t\,;\,x_o, t_o) \otimes G_{ij}(y_m, t\,;\,x_p, t_o) \qquad (4)$$

The operation $G_{ij}(y_m, -t\,;\,x_o, t_o) \otimes G_{ij}(y_m, t\,;\,x_p, t_o)$ corresponds to a matched-filter which is the correlation operator induced by the time reversal -t operation (Dorme and Fink, 1995; Fink and Prada, 2001). The optimum filter of a signal is the signal itself, $u_j(x_p, t\,;\,x_o, t_o)$. The filter output should be maximum at $t = T$ if $x_o = x_p$.

## 1.2 Construction of the whole Time Reversal Mirror

Fig. 3, step 3 considers the full experiment using a receiver/re-transmitter TRM line of $y_m$ points to re-transmit the time reversed signals recorded from an initial shot ($x_o$, $t_o$). The matched-filter for the source at $x_o$ becomes:

$$u_j(x_p, t\,;\,x_o, t_o) = \delta(t-T) \otimes \Sigma_m \{ G_{ij}(y_m, -t\,;\,x_o, t_o) \otimes G_{ij}(y_m, t\,;\,x_p, t_o) \} \qquad (5)$$

where $\Sigma_m$ denotes the contribution of all receivers $y_m$ in the "*mirror*" gallery. As shown previously, the matching summation of correlations will be maximum for $x_o = x_p$ at $t = T$ (Fig. 4).

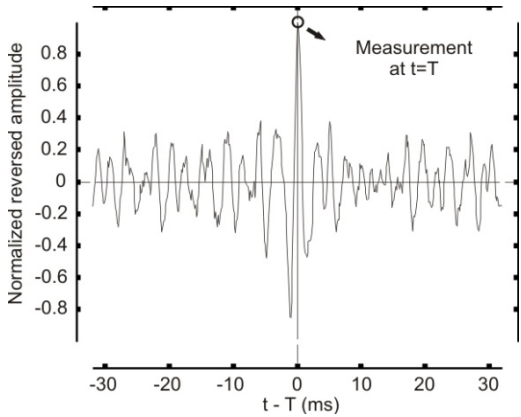
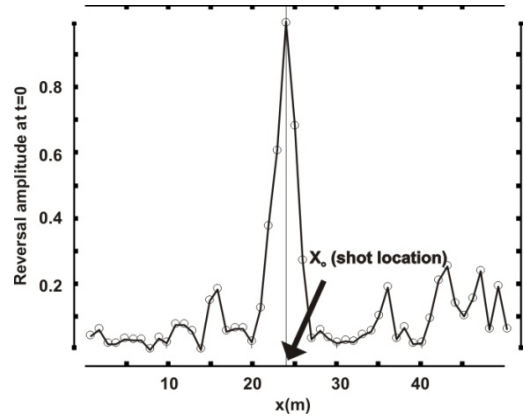

*Figure 4:* Reversed signal issued from Eq. 5. In order to better display the signal and its maximum. The time window is T sec. shifted and depicted for $x_o = x_p$. The amplitude measured is circled at $t = T$.

*Figure 5:* Ground motion amplitude $u_j(x_p, T\,;\,x_o, t_o)$ described in Eq. 5, measured for each reversed signal at $t = T$ for all possible shot locations in the source gallery. The initial shot $x_o$ is located in GAS gallery. Width at mid-height of main peak is 3 m corresponding to the refocussing coordinate $x_o = 24.0 \pm 1.5$ m.

Matching curves calculated using Eq. (5) and depicted in Fig. 4, can be obtained for each possible shot location in the "*source*" gallery. For each possible shot location, the amplitude at time *T* can be measured. Such a set of measurements gives then a reversal amplitude curve depicted in Fig. 5. This plot shows a sharp peak with maximum amplitude centred at the location of the initial shot $x_o$. The amplitudes measured at the other points along the gallery may show secondary peaks with amplitudes lower than $1/5^{th}$ the maximum amplitude observed at the shot location. These residual amplitudes may help to define the noise level of the refocusing process.

## 1.3 Hyper-focusing effect of time reversal process

The finite 50 m spatial aperture of the array and the short 0.064s temporal windows used, imply that the refocused seismic energy reaching the source location is made up of only a part of the elastic energy emitted, diffused or diffracted within the medium, as determined by applying truncated experimental Green functions. However, the sharpness of the refocusing peak coordinate allows the clear identification of the location of the initial shot (Fig. 5). An average wavelength λ in the experimental medium is estimated to be 22.5 m assuming a P-wave velocity of 4500 m/s (Sénéchal et al., 2004) and an effective frequency of 200 Hz. On the basis of Fig. 5, the observed aperture of the refocusing source estimated at mid-height of the shot location peak is about $\delta x = 3$ m. It is much tighter than $\Delta x = \lambda \times L / a = 45$ m, the expected theoretical width of the refocusing source predicted by Borcea et al (2002) in a homogeneous medium ($a = 50$ m being the physical length of the actual receiver array and $L = 100$ m the rough distance of the mirror to the source). Consequently, the effective aperture of the mirror, under the conditions of the experiment, can be rewritten $a_{eff} = \lambda \times L / \delta x$. The parameter $a_{eff}$ represents the apparent length of the receiver array. In this study, the apparent aperture $a_{eff}$ reaches approximatively 750 m if we consider the main frequency of 200 Hz. This effective aperture $a_{eff}$ is much longer than the 50 m physical aperture a. The narrow refocusing sources associated with the long apparent array aperture we obtained in this study, shows the importance of the medium heterogeneity in TRM experiments (Fink, 2001). Those results confirm the "hyper-focusing" process (Tourin et al, 2006) also named "super-resolution" process (Borcea et al, 2002).

## 2. DATA ANALYSIS

TRM analysis is carried out in the four time windows "noise", "direct field", "early coda" and "later coda" (Fig. 6). Each window contains 512 points corresponding to 0.064s duration and is high-pass filtered at 100 Hz. The shot occurred at time 0.064 s. Assuming a mean P-wave velocity of 4500 m/s (Sénéchal et al., 2004), the path length range for P-wave is approximately 100-290 m, 290-580 m and 580-860 m for the "direct field", "coda", and "later coda" windows analyzed respectively, the galleries being 100 m apart.

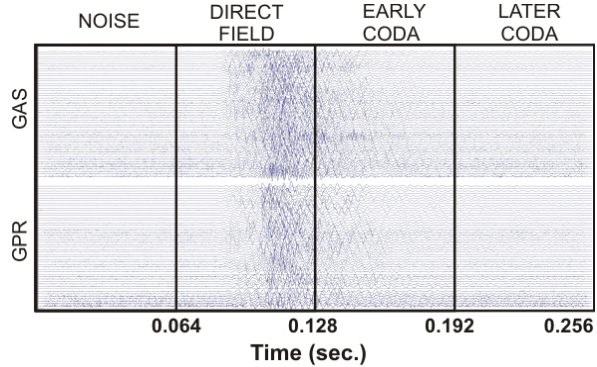

*Figure 6: Transmitted and scattered signals recorded in GPR (lower part) and GAS (upper part) galleries for two shots impacted in GAS and GPR respectively. The signals have been divided into NOISE, DIRECT FIELD, EARLY CODA and LATER CODA segments.*

The TRM analysis is shown in Fig. 7 for all the shots done in the GAS (left side) and in the GPR (right side) galleries. The vertical axes correspond to the initial coordinates of shots assuming that shot number n occurred between two positions n and n + 1, i.e. $x_n = (x_n + x_{n+1}) / 2$. Shot 15 was impacted twice inducing a step in the energy plot displayed for the GAS. The horizontal axes correspond to the reversed coordinates, i.e. related to the true sensor locations themselves, as the ones described previously in Fig. 5. In the grey scale associated with the measured reversal amplitude, the higher amplitude features, which represent a greater degree of refocusing, appear darker. The vertical white streaks correspond to failed recordings of signals at step 1 and 2 (see Fig. 3) making them impossible to reverse the shot energy.

The results show energy randomly distributed when the TRM is applied using the noise windows (Figs. 7A). Considering the other time windows (i.e. Figs. 7B, 7C, and 7D), there are clear alignments of energy, i.e. energy refocusing, along the theoretical grey line, when the TRM is applied along the "direct field", "early coda" and "later coda" windows respectively. These results strongly confirm the necessity of exciting a coherent source in order to focus the related coherent fields using the TRM technique.

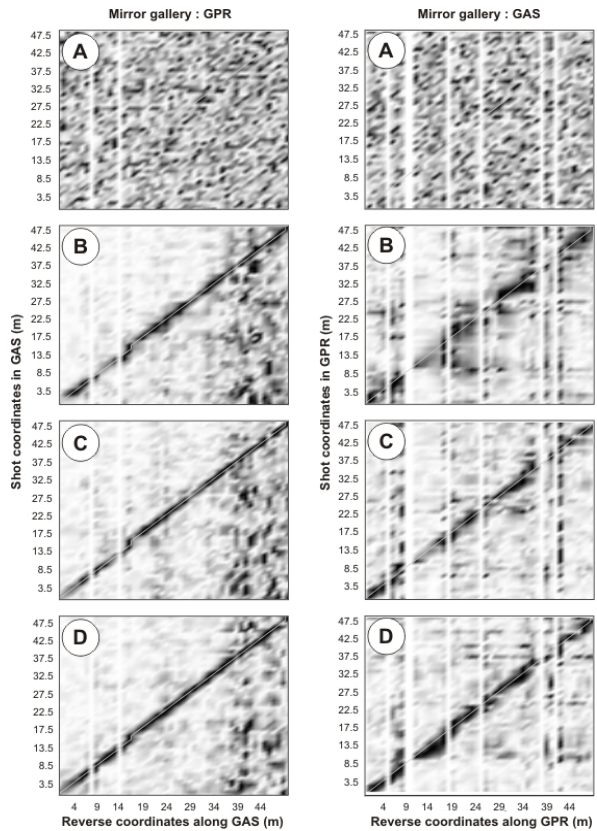

*Figure 7:* Energy location after reversal in GAS (left side) and in GPR (right side) galleries. White corresponds to minimum energy and black to maximum energy focussing. Coordinates are in meters. Windows A, B, C, and D correspond to the NOISE part of the signal (i.e. before the sledgehammer impact), the DIRECT FIELD, the EARLY CODA, and the LATER CODA respectively as reported in Fig. 6. The light grey line shows the theoretical maximum energy location.

Even if reversed energy follows the theoretical focussing line out to the most distant geophones in the GAS gallery, secondary energetic peaks above and below the ideal focussing line, occur at the deepest zone of GAS gallery with reversal coordinates in the range 36-50 m. The cause of this unfocussed energy is unclear.

Focussing appears to be enhanced with increasing coda time. It is clearly depicted by the width evolution of the refocusing ridge depicted in the boxes B to D in Fig. 7, corresponding to shots in the GAS (left side) and GPR (right side) galleries. The focussing process is emphasized when the scattering increases, an observation that is consistent with the conclusion in Fink (2001).

In a general way, the "hyper-focussing" effect is achieved with a lower quality in the GPR than in the GAS gallery since the mirror used for GPR focussing is located in the GAS corrugated rock wall gallery. The roughness of the natural rock walls in the GAS gallery makes it likely that sensor orientations may differ significantly from the directions of the source impacts on the wall. The experimental conditions in the GAS gallery make it difficult to strictly abide by the reciprocity theorem used to build Eq. (3). The rugosity of the anti-blast gallery (GAS) perturbs the ideal experimental condition resulting in a wider less resolved correlation peak for the B, C and D. This interpretation is reinforced by the thinner energy focussing line obtained when focussing is done in the GAS gallery with the mirror located on the flat concrete wall in the GPR gallery (left side in Figs. 7B, 7C and 7D).

**CONCLUSION**

The results of a TRM experiment at a mesoscopic scale confirm that elastic energy refocusing on the source location is a stable process. The existence of a source remains necessary to generate the reversal process since that source insures that the wavefield remains coherent even after diffracted and diffused waves propagate into the medium. Here coherency is defined in the sense that the wavefield is generated by a specific physical process precisely localized in space (e.g. the sledgehammer shot point). The TRM can be produced as long as field energy related to the source remains present in the time window analyzed. Depending on experimental conditions, the TRM focussing process is sensitive to the roughness of mirror construction which governs the quality of physical expression of wavefield reciprocity. It appears, finally, that the later time windows (i.e. those with longer field path duration and path length in the medium) have better refocusing quality, due to the scattered waves generated by the small scale random inhomogeneity of the medium.

**Acknowledgments**

This work was founded by the Plan Pluri-Formation "magneto-hydro-sismique" of the Université de Nice – Sophia-Antipolis, France. We thank Bart van Tiggelen and Arnaud Tourin for their encouraging comments during this work and William Luebke and Matthew Yedlin for their suggestions to improve the quality of the manuscript.